\documentclass[pra,aps]{article}
\usepackage{epsfig}

\usepackage{amssymb}
\begin{document}

\sloppy

\title{Nanosecond Dynamics of Single-Molecule Fluorescence Resonance
Energy Transfer}
\author{G.O. Ariunbold$^{1,2,5}$, G.S. Agarwal$^{1,2,3}$ Z. Wang$^{2}$
\\M.O. Scully$^{1,2,4}$ and H. Walther$^{1}$
\\
$^1$ Max-Planck-Institut f${\rm \ddot{u}}$r Quantenoptik
\\ D-85748 Garching, Germany
\\
$^2$Institute for Quantum Studies,\\ Department of Physics
\\Texas A\&M  University, College Station,
\\
TX 77843-4242, USA
\\
$^3$ Physical Research Laboratory,\\ Navrangpura, Ahmedabad-380
009, India
\\
$^4$ Department of Chemistry, Princeton University
\\Princeton, NJ 08544, USA
\\
$^5$ Theoretical Physics Laboratory,\\
National University of Mongolia
\\
210646, Ulaanbaatar, Mongolia
}
\date{\today}
%
%
\maketitle
%
\begin{abstract}

Motivated by recent experiments on photon statistics
from individual dye pairs planted on biomolecules
and coupled by fluorescence resonance energy transfer (FRET),
we show here that the FRET dynamics can be modelled by Gaussian
random processes with colored noise.
Using Monte-Carlo numerical simulations, the photon
intensity correlations from the FRET pairs are calculated,
and are turned out to be very close to those observed in experiment. The
proposed stochastic description of FRET is consistent
with existing theories for microscopic dynamics of
the biomolecule that carries the FRET coupled dye pairs.
\end{abstract}


\section{Introduction}

Recent significant advances in nano-technology make it possible to
investigate molecular dynamics and structures at single-molecule level.
Measurement of fluorescence resonant energy transfer (FRET)
between a couple of dye molecules that are attached to complementary
sites of a biomolecule like DNA or protein is particularly
useful, because the sharply distance-dependent dipole-dipole
interaction
between the dye pair can serve as a 'spectroscopic
ruler' for the biomolecule~\cite{arifoerster,arideniz,arilipman}.
FRET means a non-radiative quantum energy transfer from
a donor that is a dye which initially
absorbs light, to an acceptor which is another dye.

The FRET can be considered in the
framework of the theory of F\"{o}rster~\cite{arifoerster}.
An input laser light excites the donor,
whose one of decay channels is to migrate its
excitation energy to the acceptor via dipole-dipole
interaction. The energy transfer typically finishes within nanoseconds.
The requirements for FRET to occur efficiently are, at least, one
of the chromophores should have a sufficient quantum yield and the
donor fluorescence spectrum must overlap the acceptor absorption
spectrum.

Photon-photon correlations associated with the chromophores
contain information on the conformational distance of the
biomolecule. Such information is usually washed out in
traditional ensemble measurement, but is readily available in
single-molecular measurement. Recently, using a
Hanbury-Brown--Twiss~\cite{ariscully} time-interval apparatus,
Berglund et al.~\cite{ariberglund} have measured photon intensity
correlations for individual donor-acceptor pairs on DNA. To
interpret the experimental data, they proposed a dual FRET model.
A continuous model emphasizing overall conformational change of
the biomolecule has also been studied by several
authors~\cite{arihaas,arinarg}. However, it is important to note
that there exists non-trivial interplay between biomolecular
diffusion, which under physiological conditions can change
drastically the molecular conformation over nanoseconds, and the
quantum optical processes of a pair of FRET coupled dyes, which
are also of the time scale of nanoseconds.
In this paper, we show that the laser-induced FRET dynamics can
be modelled by the stochastic Ornstein--Uhlenbeck (OU)
process~\cite{ariwaxbook}. The underlying stochastic process is a
Gaussian random process~\cite{arikampenbook} with finite
correlation time. The density matrix equations acquire the
character of stochastic differential equations which can be
solved using well-established methods. We shall demonstrate that
the OU process is a good approximation to the FRET dynamics as
measured on biomolecules. This paper is organized as follows. In
section 2 we present our theoretical model of FRET. The idea of
modelling FRET dynamics as an OU stochastic process is further
elucidated in the context of biomolecular dynamics in solution.
In section 3 and 4 we present the results of Monte-Carlo
simulations and compare them with the experimental work.

\section{Theory}

In order to calculate the photon-photon correlation functions, we
adopt a master equation approach. To derive the master equation
for the system with two molecules coupled by FRET, we assume that
the coherence time is much shorter than the time scale of
experiments.
We further note that a FRET coupled system is not a cascaded
quantum system~\cite{arigardinerbook}, i.e., to obtain an equation
for only donor variables, after tracing over the acceptor
variables from the general master equation, is impossible.
The master equation for the density operator $\hat{\varrho}$, for
the donor-acceptor pair couples four possible states
$|q\rangle_1\otimes |p\rangle_2$ here $q=0,1$ ($p=0,1$) stands
for the ground and excited states of the donor
(acceptor)~\cite{ariberglund}:
\begin{equation}  \label{ariopeqn}
\frac{d\hat{\varrho}}{dt}=\sum_{m=1}^5\Gamma_m(t) \left\{
\hat{\Gamma}_m\hat{\varrho}\hat{\Gamma}_m^\dagger
-\frac{1}{2}\left(\hat{\Gamma}_m^\dagger\hat{\Gamma}_m\hat{\varrho}
+\hat{\varrho}\hat{\Gamma}_m^\dagger\hat{\Gamma}_m
\right)\right\}
\end{equation}
where $\Gamma_m, m=1-5$ are rate coefficients for the dominant
processes: spontaneous emissions of the donor and acceptor
are described by the quantum jump operators
$\hat{\Gamma}_1=\hat{\sigma}_1\otimes \hat{1}_2$ and
$\hat{\Gamma}_2=\hat{1}_1\otimes \hat{\sigma}_2$;
laser excitations of
the donor and acceptor by
$\hat{\Gamma}_3=\hat{\sigma}_1^+\otimes \hat{1}_2$ and
$\hat{\Gamma}_4=\hat{1}_1\otimes \hat{\sigma}_2^+$;
and transfer of energy from the donor to the
acceptor by
$\hat{\Gamma}_5=\hat{\sigma}_1\otimes \hat{\sigma}_2^+$.
Here $\hat{\sigma}_i$ ($\hat{\sigma}_i^+$) is the lowering (raising)
operator between the excited and the ground states for
the $i$th dye molecule ($i=1$ for the donor and 2 for the acceptor).
The master
equation (\ref{ariopeqn}) ignores all coherence effects as they
play no important role in FRET. The conditional photon count
probability of
the donor-acceptor system within a time delay $\tau$ can be
represented as a normally ordered correlation
function~\cite{arigardinerbook}
\begin{equation}  \label{aricorr}
\langle :\hat{I}_i(t+\tau)\hat{I}_j(t):\rangle = {\rm
Tr}_{i,j}\left\{\hat{\sigma}_{i}^+\hat{\sigma}_{i}\hat{V}(\tau)\left\{
\hat{\sigma}_{j}\hat{\varrho}(t)\hat{\sigma}_{j}^+\right\}\right\}
\end{equation}
here $\hat{\varrho}(t)$ is
the stationary operator solution of Eq.~(\ref{ariopeqn}), and
$\hat{V}(\tau)$ is the evolution operator of the whole system
satisfying $\hat{V}(\tau)\hat{\varrho}(t)=\hat{\varrho}(t+\tau)$.

FRET measurements provide us information about dipole-dipole
coupling, which varies as $1/{r^3}$~\cite{ariagarwalbook}. The
FRET rate coefficient is given by
\begin{equation}  \label{arifretrate}
\Gamma_5(t)=\Gamma_1\left(\frac{R_0}{r(t)} \right)^6
\end{equation}
where $R_0$ is the F\"{o}rster radius, $r(t)=|{\bf r}_1(t)-{\bf r}_2(t)|$ is the distance
between two chromophores. Considering intra-chain diffusion of a biomolecule, e.g.,
a protein molecule that carries the donor and the acceptor at
a couple of complementary sites, the displacement of the donor-acceptor distance from its fixed (or initial)
value, $x(t) = r(t)- r(0)$, can be modelled as
a Brownian motion in a harmonic potential. Since a biomolecule which consists of a
large number of atoms and molecules, we essentially deal with the over-damped regime, where
the Brownian motion is described by the Langevin
equation~\cite{arikampenbook}
\begin{equation}
\frac{\partial x(t)}{\partial t} = - \lambda x(t) + \lambda f_x(t)
\end{equation}
where $\lambda = \omega^2/\eta$ with $\omega$ being the frequency of
harmonic oscillator, and $\eta$ the friction coefficient. The random
force $f_x(t)$ is a pure Gaussian characterized by $\langle f_x(t) \rangle
=0$ and $\langle f_x(t)f_x(t+t') \rangle = 2\theta \delta (t-t')$.
The solution of the Langevin equation, $x(t)$, resembles a white noise:
its equilibrium distribution is
\begin{equation}
\rho^{eq}(x) =\frac{1}{\sqrt{2\pi \theta\lambda}} {\rm exp}\left(-\frac{x^2}{2\theta\lambda}\right).
\end{equation}
and the correlation is
\begin{equation}  \label{ariexpcorr0}
\langle x(t)x(t') \rangle = \theta\lambda e^{-\lambda |t-t'|}.
\end{equation}

Now turn to the FRET rate Eq.~(\ref{arifretrate}), which can be rewritten as
$\Gamma_5(t)=  \Gamma_5(0)/[1+x(t)/r(0)]^6$ with $\Gamma_5(0) = \Gamma_1 [R_0/r(0)]^6$ corresponding to the FRET
rate at some fixed interdye distance. If the displacement $x$ is small as compared to $r(0)$, then we have
\begin{equation} \label{arirelation}
\delta \Gamma_5(t) \sim x(t).
\end{equation}
This is a possible interplay between the diffusion process and the FRET dynamics, although
neither they are completely uncorrelated nor fully correlated.
It turns out that a variance (i.e., `noise') of the FRET rate
has the statistical signature of a white noise. A sign such as 'noise of the noise' usually leads to
colored noise in the OU stochastic dynamics.
Eq.(\ref{arirelation}) is a linearized approximate relation.
Although, some dynamic details can not be seen for large
variations, analytically, the exact numerical simulation may be
done. This rather involved and so we have tried to produce
reasonable results by retaining the leading term. This captures
the physics reasonably well.
Note that the OU stochastic process is stationary. Thus
Eq.(\ref{ariopeqn}) becomes a stochastic differential equation as
$\Gamma_5(t)=\Gamma_5(0)+\xi(t)$. We assume that $\xi(t)$ is a
colored noise, which is described by the Langevin
equation~\cite{aricoffeybook}
\begin{equation}  \label{arieqncol}
\frac{d}{dt}\xi=-\lambda \xi + \lambda \eta(t)
\end{equation}
where time averages of white noise should be $\overline{\eta(t)}=0$ and $\overline{\eta(t)\eta(t')}=2D\delta(t-t')$.
As well known the Eq.~(\ref{arieqncol}), yields the steady state
correlation function
\begin{equation}    \label{ariexpcorr}
\left\{ \overline{\xi(t)\xi(t')} \right \} = D \lambda {\rm
e}^{-\lambda |t-t'|}
\end{equation}
with $\overline{\xi(t)}=0$ and $\{...\}$ denotes the stochastic
average over the initial conditions~\cite{arivemuri}. A parameter
$D$ might be proportional to the diffusion coefficient $\theta$
and $\lambda$ is the same in Eq.(\ref{ariexpcorr0}) and
Eq.(\ref{ariexpcorr}). The stochastic differential equation
Eq.(\ref{ariopeqn}) will be solved using Monte-Carlo numerical
simulations. A Box-Mueller algorithm and the Euler-Maruyama method
have been used to realize the colored noise. Moreover, by virtue
of an integral algorithm developed in~\cite{arivemuri}, we have
verified that the Monte-Carlo generated correlation fits
perfectly to its analytical expression Eq.~(\ref{ariexpcorr}). To
achieve this, a stochastic averaging over as large as $1000$
realizations is essential~\cite{ariagarwal}.
It must be borne in mind that the FRET rate is always positive
which is done by keeping a background constant value
$\Gamma_5(0)$. However, some large negative random numbers have to
be omitted. To prove that these omitted random numbers do not
play important role, we have also calculated the correlation for
$\Gamma_5(t)$ which decreases exponentially as given in
Eq.(\ref{ariexpcorr}). Namely, we assume that $\Gamma_H\ge
\Gamma_5(t)\ge \Gamma_L$ where the FRET rate is finite.
$\Gamma_H$ ($\Gamma_L$) corresponds to minimum (maximum)
inter-dye distance due to continues intrachain diffusion of the
protein molecule. The quantities are determined by contour length
and bending rigidity of the protein. In our case, we have assumed
that the FRET does not occur between distant dyes, so that
$\Gamma_L=0$. We take also $\Gamma_H=max(\Gamma_5(t))$ as a
maximum value of the generated random numbers.

\section{Results and discussions}

Using Monte-Carlo simulations we have calculated correlation functions
as defined in Eq.~(\ref{aricorr}). In Fig.~1 we present the results
for the normalized correlations $g^{(2)}_{ij}(\tau)$ defined by
\begin{equation}
g^{(2)}_{ij}(\tau)=\frac{\langle :
\hat{I}_i(t+\tau)\hat{I}_j(t):\rangle}
{\langle : \hat{I}_i(t) \hat{I}_j(t) :\rangle}.
\end{equation}
Since only two parameters, $D$ and $\lambda$, are needed to
completely specify an OU process, their determination
would be a desired contact between theory and experiment.
As we see in Fig.~1, the normalized correlation functions are
very similar to those observed in the
experiment\cite{ariberglund}. For the data shown in Fig.~1 the
correlation time is taken to be $\tau_c=1/\lambda=7$. In this
case we take the parameter $\Gamma_5(0)$ to be $0.65$ in order to
ensure that the average FRET coefficient to be around $1$. The
FRET coefficients fluctuate between $\Gamma_L=0$ and $\Gamma_H=5$.
The estimated F\"{o}rster radius, for instance, for the TMR-Cy5
dye pair that is frequently used in biomolecular measurement, is
about $53$ \AA~\cite{arideniz}. Given the calculated average FRET
coefficient of $1$, the average distance would be approximately
$50$ \AA. Intensity autocorrelations show the typical quantum
feature of antibunching that is characteristic of emissions from
individual dye molecules. Following the initial photon
antibunching, photon bunching appears in the acceptor
autocorrelation function. A sufficient stochastic deviation from
its equilibrium distribution of the FRET rate is the hallmark of
the generation of photon bunching, which is a tendency for
clustered emissions. For large $\tau$ the normalized correlation
functions go to unity indicating uncorrelated emissions. The
appearance of the photon bunching in the donor autocorrelation
was first predicted theoretically by Haas and
Steinberg~\cite{arihaas}. A pronounced antibunching associated
with a photon blockade effect and a photon bunching in the cross
talk of the acceptor-donor pair, for short time, has also been
discussed in~\cite{ariberglund}. We also notice in the
experimental results that over longer times, photons emitted by
the donor are becoming correlated with photons emitted by the
acceptor and vice versa. It is also worth noting that because of
off-resonant excitation of the acceptor, the corresponding rate
for the acceptor is as small as $f \Gamma_3$ with $f=0.1$.
Otherwise, the acceptor would have already been excited by the
laser, and FRET may not occur. For the same reason the laser
excitation rate should not be too large. The bunching signal can
be described approximately by an exponential $y=1+C e^{-\lambda_0
\tau}$, with the fitting parameter $\lambda_0$ proportional to
$\lambda$. The tail of exponential decay, especially in the
acceptor autocorrelation function is determined by the
correlation time $\tau_c$ of the OU process. It is obvious that a
larger value of the correlation time results in a longer tail.
While being supported by the simulation results (Fig.~1), this
point can be made clearer by assuming that the intra-chain
diffusion of the biomolecule that carries the dyes, and thereby
fluctuation of the donor-acceptor distance is much slower than
the quantum optical process of the donor and acceptor. The
population on the excited states then adiabatically follows the
slowly varying $\Gamma_5(t)$, so that the intensity of the
acceptor can be approximated as\cite{arinarg}
\begin{equation} \label{ariintensity}
I_2(t) = \Gamma_3\left( f+\frac{ \Gamma_5(r(t)) }{\Gamma_5(r(t))+\Gamma_1}
\right).
\end{equation}
According to this adiabatic approximation in longer time delay $\tau$,
the intensity correlation is an exponential~\cite{arinarg}
\begin{equation} \label{ariadiabat}
g_{22}^{(2)}(\tau) \simeq 1 + \frac{(I_H -I_L)^2}{(I_H + I_L)^2}
e^{-2t/\tau}
\end{equation}
where we have assumed that the dual FRET rates result in a high
value and a low value of the emission intensity, $I_H$ and $I_L$,
respectively. Using $\Gamma_H=5$ and $\Gamma_L=0$ obtained in the
calculation for Fig.~1, we find that $I_H=0.93$ and $I_L=0.1$
from Eq.(\ref{ariintensity}). The correlation function under the
adiabatic approximation given by Eq.(\ref{ariadiabat}) has been
also plotted, see the dotted curve in Fig.~1.

\section{Conclusions}
%
We have shown how the FRET process as measured for biomolecules
in solution can be modelled using the Ornstein-Uhlenbeck
stochastic theory. This theory predicts the fluorescence intensity
correlations from the FRET coupled dye pair, which are very
similar to those observed in recent experiments. An analytic
study based on the local linearization procedure also shows that
it is consistent with existing theory for microscopic dynamics of
the biomolecule that carries the FRET coupled dye pairs. The
second-order intensity correlation functions for a FRET coupled
dye pair, are largely determined by only a few statistical
parameters of the FRET dynamics. It is found that the stochastic
OU description helps elucidate the underlying mechanism for the
experimentally observed fluorescence correlations that typically
exhibit exponential decay over nanosecond timescale.

{\bf{ Acknowledgement.}} This work was supported by ONR grant
N00014-03-1-0639 and N00014-02-1-0741; Welch Foundation grant A-1261;
AFRL grant F30602-01-1-0594 and TITF. One of us, (G.O.A.) also
acknowledges the support of the Humboldt Foundation Fellowship.


%
\begin{figure}[!ht]
\begin{center}
\includegraphics[width=68mm]{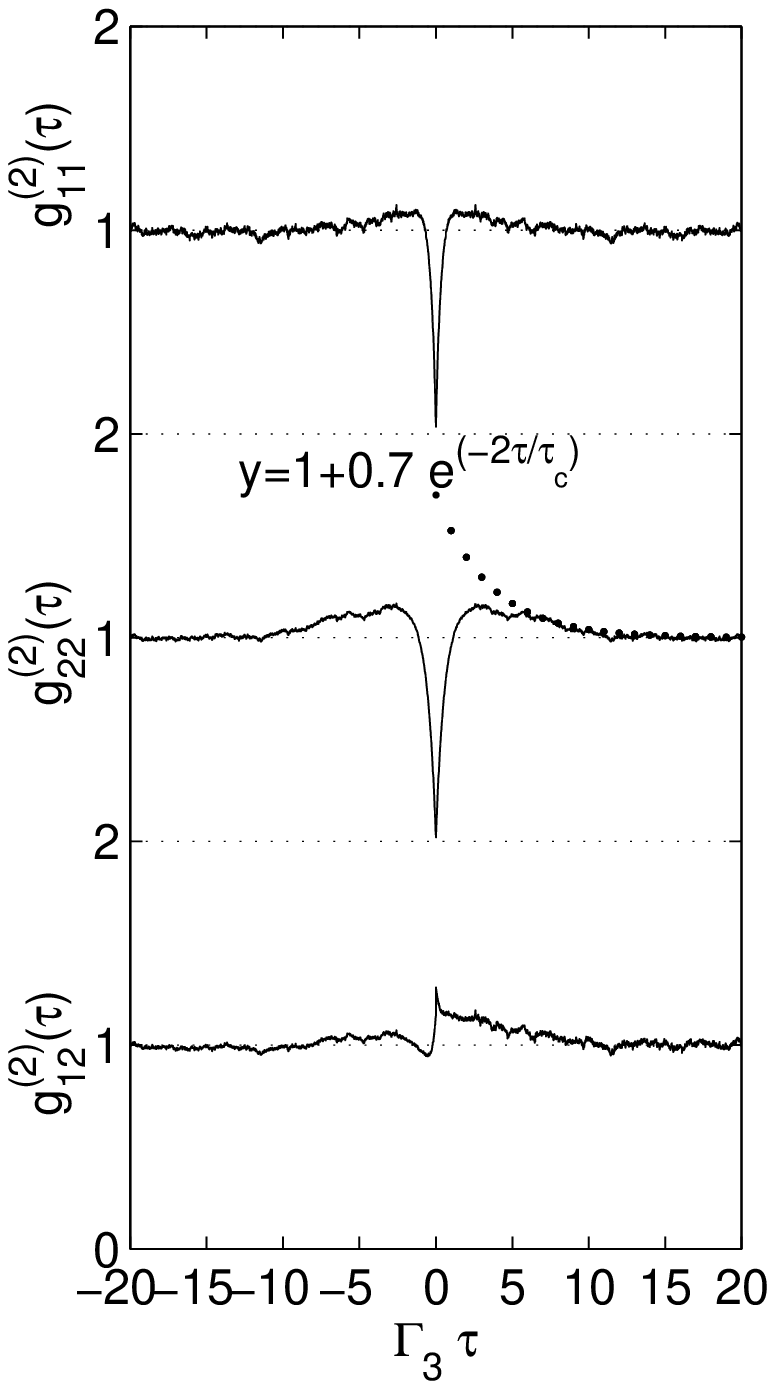}
\label{arifig1} \caption{Normalized photon-photon correlation
functions corresponding to donor-donor, acceptor-acceptor and
donor-acceptor emissions. The parameters of the model
(\ref{ariopeqn}) are chosen as $\Gamma_{1,2,3}=1, \Gamma_4=0.1,
\Gamma_5(0)=0.65, \ (\overline{\Gamma_5(t)} \simeq 1, \ 0 \le
\Gamma_5(t)\le 5)$. The noise parameters are taken as
$\tau_c=1/\lambda=7$ and $D=7$. The number of numerical
realizations and a time step are taken to be $N=1000$ and
$dt=0.01$, respectively.}
\end{center}
\end{figure}
%

\end{document}